\documentclass{stacs_proc}

\usepackage{definitions}

\begin{document}

\title[Order-Invariant MSO is Stronger than Counting MSO]
  {Order-Invariant MSO is Stronger than Counting MSO\linebreak[3]
    in the Finite} 

\author[lab1]{T.~Ganzow}{Tobias Ganzow}
\address[lab1]{Mathematische Grundlagen der
    Informatik, RWTH Aachen, Germany}  
\email{ganzow@logic.rwth-aachen.de}  

\author[lab2]{S.~Rubin}{Sasha Rubin}
\address[lab2]{Department of Computer Science, University
    of Auckland, New Zealand}	
\email{rubin@cs.auckland.ac.nz}  


\begin{abstract}
  We compare the expressiveness of two extensions of monadic
  second-order logic (MSO) over the class of finite structures. The
  first, counting monadic second-order logic (CMSO), extends MSO with
  first-order modulo-counting quantifiers, allowing the expression of
  queries like ``the number of elements in the structure is even''.
  The second extension allows the use of an additional binary
  predicate, not contained in the signature of the queried structure,
  that must be interpreted as an arbitrary linear order on its
  universe, obtaining order-invariant MSO.

  While it is straightforward that every CMSO formula can be
  translated into an equivalent order-invariant MSO formula, the
  converse had not yet been settled. Courcelle showed that for
  restricted classes of structures both order-invariant MSO and CMSO
  are equally expressive, but conjectured that, in general,
  order-invariant MSO is stronger than CMSO.

  We affirm this conjecture by presenting a class of structures that
  is order-invariantly definable in MSO but not definable in CMSO.
\end{abstract}

\maketitle

\stacsheading{2008}{313-324}{Bordeaux}
\firstpageno{313}

\section{Introduction}
Linear orders play an important role in descriptive complexity theory
since certain results relating the expressive power of logics to
complexity classes, \eg, the Immerman-Vardi Theorem that \LFP captures
\CCP, only hold for classes of linearly ordered structures. Usually,
the order only serves to systematically access all elements of the
structure, and consequently to encode the configurations of a
step-wise advancing computation of a Turing machine by tuples of
elements of the structure. In these situations we do not actually want
to make statements about the properties of the order, but merely want
to have an arbitrary linear order available to express the respective
coding techniques.

Furthermore, when actually working with finite structures in an
algorithmic context, \eg, when evaluating queries in a relational
database, we are in fact working on an implicitly ordered structure
since, although relations in a database are modelled as \emph{sets} of
tuples, the relations are nevertheless stored as \emph{ordered
  sequences} of tuples in memory or on a disk. As this linear order is
always available (though, as in the case of databases, it is
implementation-dependent and may even change over time as tuples are
inserted or deleted), we could allow queries to make use of an
additional binary predicate that is interpreted as a linear order on
the universe of the structure, but require the outcome of the query
not to depend on the actual ordering, but to be
\emph{order-invariant}. Precisely, given a $\tau$-structure $\StrA$,
we allow queries built over an expanded vocabulary $\tau \dunion
\{<\}$, and say that a query $\phi$ is \emph{order-invariant} if
$(\StrA,<_1) \models \phi \ \Longleftrightarrow\ (\StrA,<_2) \models
\phi$ for all possible relations $<_1$ and $<_2$ linearly ordering
$A$.

Using \EF-games for MSO, one can see that MSO on sets (\ie, structures
over an empty vocabulary) is too weak to express that the universe
contains an even number of elements.  However, this is possible if the
universe is linearly ordered: simply use the MSO sentence stating that
the maximal element should be contained in the set of elements on even
positions in the ordering. Obviously, such a sentence is
order-invariant since rearranging the elements does not affect its
truth value. Gurevich uses this observation to show that the property
of Boolean algebras having an even number of atoms, although not
definable in FO, is order-invariantly definable in FO (simulating the
necessary MSO-quantification over sets of atoms by FO-quantification
over the elements of the Boolean algebra).

If we explicitly add modulo-counting to MSO, \eg, via modulo-counting
first-order quantifiers such as ``there exists an even number of
elements $x$ such that \dots'', we obtain \emph{counting monadic
  second-order logic} (CMSO), and the question naturally arises as to
whether there are properties not expressible in CMSO that can be
expressed order-invariantly in MSO.

In fact, a second separation example due to Otto gives a hint in that
direction. The class of structures presented
in~\cite{Otto:epsilon-invariance} even separates order-invariant FO
from FO extended by arbitrary unary generalised quantifiers, \ie,
especially modulo-counting quantifiers, and exploits the idea of
\qq{hiding} a part of the structure such that it is only meaningfully
usable for queries in presence of a linear order (or, as actually
proven in the paper, in presence of an arbitrary choice function).

The expressiveness of CMSO has been studied, \eg, in
\cite{Courcelle:MSO-I}, where it is mainly compared to \MSO, and in
\cite{Courcelle:MSO-X} it is shown that, on the class of forests,
order-invariant MSO is no more expressive than CMSO. As pointed out in
\cite{BenSeg:TameStructs}, this can be generalised using results in
\cite{Lapoire98} to classes of structures of bounded tree-width. But
still, this left open Courcelle's conjecture: that order-invariant MSO
is strictly stronger than CMSO for general graphs
\cite[Conjecture~7.3]{Courcelle:MSO-X}.

In this paper, we present a suitable characterisation of
CMSO-definability in terms of an \EF game, and later, as the main
contribution, we present a separating example showing that a special
class of graphs is indeed definable by an order-invariant MSO sentence
but not by a counting MSO sentence.

\section{Preliminaries}

Throughout the paper $\setN$ denotes the set of non-negative integers
and $\setN^+ := \setN-\{0\}$. Given a non-empty finite set $M =
\{\seq[k]m\} \finsubseteq \setN^+$, let $\lcm(M) := \lcm(\seq[k]m)$
denote the least common multiple of all elements in $M$; additionally,
we define $\lcm(\emptyset) = 1$. For sets $X$ and $Y$ as well as $M$
as before, we abbreviate that $|X| \equiv |Y| \pmod m$ for all $m \in
M$ by using the shorthand $|X| \equiv |Y| \pmod M$.

We restrict our attention to finite $\tau$-structures with a nonempty
universe over a countable relational vocabulary $\tau$, possibly with
constants, and we will mainly deal with monadic second-order logic and
some of its extensions. For more details concerning finite model
theory, we refer to \cite{EF:FMT} or \cite{Libkin:FMT}.

When comparing the expressiveness of two logics $\Logic$ and
$\Logic'$, we say that \emph{$\Logic'$ is at least as expressive as
  $\Logic$}, denoted $\Logic \subseteq \Logic'$, if for every
$\phi\in\Logic[\tau]$ there exists a $\phi'\in\Logic'[\tau]$ such that
$\Mod(\phi) = \Mod(\phi')$, where $\Mod(\phi)$ denotes the class of
all finite $\tau$-structures satisfying $\phi$.

\subsection{Counting MSO}
The notion of (modulo-)counting monadic second-order logic (\CMSO) can
be introduced in two different, but nonetheless equivalent, ways. The
first view of \CMSO is via an extension of \MSO by modulo-counting
first-order quantifiers.  

\begin{definition}
  Let $\tau$ be a signature and $M \subseteq \setN^+$ a set of moduli,
  then
  \begin{itemize}
  \item every formula $\phi \in \MSO[\tau]$ is also a formula in
    $\CMSO^{(M)}[\tau]$, and
  \item if $\phi(x) \in \CMSO^{(M)}[\tau]$ and $m\in M$, then
    $\exists^{(m)}x. \phi(x) \in \CMSO^{(M)}[\tau]$.
  \end{itemize}
  If we do not restrict the set of modulo-counting quantifiers being
  used, we get the full language $\CMSO[\tau] =
  \CMSO^{(\setN^+)}[\tau]$.  The semantics of \MSO formulae is as
  expected, and we have $\StrA \models \exists^{(m)}x.\phi(x)$ if and
  only if $\card{\{a\in A : \StrA\models\phi(a)\}} \equiv 0 \pmod m$.
  The quantifier rank $\qr(\psi)$ of a $\CMSO[\tau]$ formula $\psi$ is
  defined as for \MSO-formulae with the additional rule that
  $\qr\big(\exists^{(m)}x.\phi(x)\big) = 1+\qr(\phi)$, \ie, we do not
  distinguish between different kinds of quantifiers.
\end{definition}

In this paper we use an alternative but equivalent definition of
\CMSO, namely the extension of the \MSO language by monadic
second-order predicates $C^{(m)}$ which hold true of a set $X$ if and
only if $|X| \equiv 0 \pmod m$. As in the definition above, formulae
of the fragment $\CMSO^{(M)}[\tau]$ may only use predicates $C^{(m)}$
where $m \in M$. The back-and-forth translation can be carried out
along the following equivalences which increase the quantifier rank by
at most one in each step:
\begin{align*}
  \exists^{(m)}x.\phi(x)\ &\equiv\ \exists X(C^{(m)}(X) \land
  \forall x(Xx \liff \phi(x)))\quad\text{and}\\
  C^{(m)}(X) &\equiv\ \exists^{(m)}x. Xx\,.
\end{align*}

Furthermore, the introduction of additional predicates $C^{(m,r)}$
(or, equivalently, additional modulo-counting quantifiers
$\exists^{(m,r)}$) stating for a set $X$ that $|X|\equiv r \pmod m$
does not increase the expressive power since they can be simulated as
follows (with only a constant increase of quantifier rank):
\begin{align*}
  C^{(m,r)}(X)\ & \equiv\ \exists X_0 (\text{\qq{$X_0 \subseteq X$}}
  \land \text{\qq{$|X_0| = r$}} \land \text{\qq{$C^{(m)}(X\setminus
      X_0)$}})\,,
\end{align*}
where all subformulae are easily expressible in \MSO.

Later, we will introduce an \EF game capturing the expressiveness of
\CMSO with this extended set of second-order predicates.

\subsection{Order-invariance}
Let $\tau$ be a relational vocabulary and $\phi \in \MSO[\tau \dunion
\{<\}]$, \ie, $\phi$ may contain an additional relation symbol $<$.
Then $\phi$ is called \emph{order-invariant on a class $\ClsC$ of
  $\tau$-structures} if, and only if, $(\StrA,<_1) \models \phi\
\Longleftrightarrow\ (\StrA,<_2) \models \phi$ for all $\StrA \in
\ClsC$ and all linear orders $<_1$ and $<_2$ on $A$.

Although, in general, it is undecidable whether a given \MSO-formula
is order-invariant in the finite, we will speak of the
\emph{order-invariant fragment of} \MSO, denoted by $\iMSO{<}$, that
contains all formulae that are order-invariant on the class of all
finite structures.

It is an easy observation that every $\CMSO$ formula is equivalent
over the class of all finite structures to an order-invariant \MSO
formula by translating counting quantifiers in the following way:
\begin{align*}
  \cntexists{q}x.\phi(x) &\ :=\
  \exists X \exists X_0 \ldots \exists X_{q-1}\\
  & \qquad\left( \begin{aligned}
      & \forall x \left(Xx \liff \phi(x)\right)
      \land\ \text{``$\{X_0,\ldots,X_{q-1}\}$ is a partition of $X$''\ }\\
      \land\ & \exists x\big(X_0x \land \forall y(Xy \limp x \leq y)\big)
      \land\ \exists x\big(X_{q-1}x \land \forall y(Xy \limp x \geq y)\big)\\
      \land\ & \forall x \forall y
      \left(S_{\phi,<}(x,y) \limp \left(
          \bigland_{i=0}^{q-1}X_i x \liff X_{i+1\!\!\pmod q} y
        \right)\right)
    \end{aligned}\right)
\end{align*}
where $S_{\phi,<}$ defines the successor relation induced by an
arbitrary order $<$ on the universe of the structure restricted to the
set $X$ of elements for which $\phi$ holds.

Note that the quantifier rank of the translated formula is not
constant but bounded by the parameter in the counting quantifier.

\section{An \EF game for CMSO}

The \EF game capturing expressiveness of \MSO parameterised by the
quantifier-rank (cf.\ \cite{EF:FMT,Libkin:FMT}) can be naturally
extended to a game capturing the expressiveness of \CMSO parameterised
by the quantifier rank and the set of moduli being used in the
cardinality predicates or counting quantifiers.

Viewing \CMSO as \MSO with additional quantifiers
$\exists^{(m)}x.\phi(x)$ for all $m$ in a fixed set $M$ leads to a new
type of move described, \eg, in the context of extending \FO by
modulo-counting quantifiers in \cite{Nurmonen:mod-quant}. Since a
modulo-counting quantifier actually combines notions of a first-order
and a monadic second-order quantifier in the sense that it makes a
statement about the cardinality of a certain \emph{set} of elements,
but on the other hand, it behaves like a first-order quantifier
binding an \emph{element} variable and making a statement about that
particular element, the move capturing modulo-counting quantification
consists of two phases.  First, Spoiler and Duplicator select sets of
elements~$S$ and~$D$ in the structures such that $|S| \equiv |D| \pmod
M$, and in the second phase, Spoiler and Duplicator select elements
$a$ and $b$ such that $a\in S$ if and only if $b\in D$. After the
move, reflecting the first-order nature of the quantifier, only the
two selected elements $a$ and $b$ are remembered and contribute to the
next position in the game, whereas the information about the chosen
sets is discarded.

We prefer viewing CMSO via second-order cardinality predicates,
yielding an \EF game that allows a much clearer description of winning
strategies. Since we do not have additional quantifiers, we have
exactly the same types of moves as in the \EF game for \MSO, and we
merely modify the winning condition to take the new predicates into
account.

Towards this end, we first introduce a suitable concept of partial
isomorphisms between structures.

\begin{definition}
  With any structure $\StrA$ and any set $M\finsubseteq\setN^+$ we
  associate the (first-order) power set structure $\StrA^M :=
  \big(\Pot{A},(C^{(m,r)})_{\substack{m\in M\\0\leq r < m}}\big)$,
  where the predicates $C^{(m,r)}$ are interpreted in the obvious way.
  (Note that first-order predicates in the power set structure
  $\StrA^M$ naturally correspond to second-order predicates in
  $\StrA$.)

  Let $\StrA$ and $\StrB$ be $\tau$-structures, and let $M
  \finsubseteq \setN^+$ be a fixed set of moduli. Then the mapping
  $(\seq[s]{A},\seq[t]{a}) \mapsto (\seq[s]{B},\seq[t]{b})$ is called
  a \emph{twofold partial isomorphism between $\StrA$ and $\StrB$ with
    respect to $M$} if
  \begin{enumerate}[\indent\!\!\!(i)]
  \item $(\seq[t]a) \mapsto (\seq[t]b)$ is a partial isomorphism
    between $(\StrA,\seq[s]{A})$ and $(\StrB,\seq[s]{B})$ and
  \item $(\seq[s]A) \mapsto (\seq[s]B)$ is a partial isomorphism
    between $\StrA^M$ and $\StrB^M$.
  \end{enumerate}
\end{definition}

We propose the following \EF game to capture the expressiveness of 
CMSO where the use of moduli is restricted to a (finite) set $M$ and 
formulae of quantifier rank at most $r$.

\begin{definition}[\EF game for CMSO]
  Let $M \finsubseteq \setN^+$ and $r \in \setN$. The $r$-round (mod
  $M$) \EF game $\Game_r^M(\StrA,\StrB)$ is played by Spoiler and
  Duplicator on $\tau$-structures $\StrA$ and $\StrB$. In each turn,
  Spoiler can choose between the following types of moves:
  \begin{itemize}
  \item \emph{point move:} Spoiler selects an element in one of the
    structures, and Duplicator answers by selecting an element in the
    other structure.
  \item \emph{set move:} Spoiler selects a set of elements $X$ in one
    of the structures, and Duplicator responds by choosing a set
    of elements $Y$ in the other structure.
  \end{itemize}
  After $r = s+t$ rounds, when the players have chosen sets $\seq[s]A$
  and $\seq[s]B$ as well as elements $\seq[t]a$ and $\seq[t]b$ in an
  arbitrary order, Duplicator wins the game if, and only if,
  $(\seq[s]{A},\seq[t]{a}) \mapsto (\seq[s]{B},\seq[t]{b})$ is a
  twofold partial isomorphism between $\StrA$ and $\StrB$ with respect
  to $M$.
\end{definition}

First note that, although Duplicator is required to answer a set move
$X$ by a set $Y$ such that $|X| \equiv |Y| \pmod M$ in order to win,
we do not have to make this explicit in the rules of the moves since
these cardinality constraints are already imposed by the winning
condition ($X$ and $Y$ would not define a twofold partial isomorphism
if they did not satisfy the same cardinality predicates).
Furthermore, for $M = \emptyset$ or $M=\{1\}$, the resulting game
$\Game_r^M(\StrA,\StrB)$ corresponds exactly to the usual \EF game for
\MSO.

\begin{theorem}
  Let $\StrA$ and $\StrB$ be $\tau$-structures, $r\in\setN$, and $M
  \finsubseteq \setN$. Then the following are equivalent:
  \begin{enumerate}[(i)]
    \setlength{\itemsep}{1ex plus0.5ex minus0.5ex}
  \item $\StrA \equiv_r^M \StrB$, \ie, $\StrA \models \phi$ if and
    only if $\StrB \models \phi$ for all $\phi \in \CMSO^{(M)}[\tau]$
    with $\qr(\phi) \leq r$.
  \item Duplicator has a winning strategy in the $r$-round (mod $M$)
    \EF game $\Game_r^M(\StrA,\StrB)$.\qed
  \end{enumerate}
\end{theorem}

To prove non-definability results, we can make use of the following
standard argument.

\begin{proposition}
  A class $\ClsC$ of $\tau$-structures is not definable in CMSO if,
  for every $r \in \setN$ and every $M \finsubseteq \setN^+$, there
  are $\tau$-structures $\StrA_{M,r}$ and $\StrB_{M,r}$ such that
  $\StrA_{M,r} \in \ClsC$, $\StrB_{M,r} \not\in \ClsC$, and
  $\StrA_{M,r} \equiv_r^M \StrB_{M,r}$.
\end{proposition}

The following lemma, stating that the CMSO-theory of disjoint unions
can be deduced from the CMSO-theories of the components, can either be
proved, as carried out in \cite[Lemma~4.5]{Courcelle:MSO-I}, by giving
an effective translation of sentences talking about the disjoint union
of two structures into a Boolean combination of sentences each talking
about the individual structures, or by using a game-oriented view
showing that winning strategies for Duplicator in the games on two
pairs of structures can be combined into a winning strategy on the
pair of disjoint unions of the structures.

\begin{lemma}\label{lem:cmso-disjoint-union}
  Let $\StrA_1, \StrA_2, \StrB_1,$ and $\StrB_2$ be $\tau$-structures
  such that $\StrA_1 \equiv_r^M \StrB_1$ and $\StrA_2 \equiv_r^M
  \StrB_2$. Then $\StrA_1 \dunion \StrA_2 \equiv_r^M \StrB_1 \dunion
  \StrB_2$.
\end{lemma}
\begin{proof}
  Consider the game on $\StrA := \StrA_1 \dunion \StrA_2$ and $\StrB
  := \StrB_1 \dunion \StrB_2$. A Spoiler's point move in $\StrA$
  (resp., in $\StrB$) is answered by Duplicator according to her
  winning strategy in either $\Game^M_r(\StrA_1,\StrB_1)$ or
  $\Game^M_r(\StrA_2,\StrB_2)$. A set move $S \subseteq A$ (analogous
  for $S \subseteq B$) is decomposed into two subsets $S_1 := S \cap
  A_1$ and $S_2 := S \cap A_2$, and is answered by Duplicator by the
  set $D := D_1 \cup D_2$ consisting of the sets $D_1$ and $D_2$
  chosen according to her winning strategies as responses to $S_1$ and
  $S_2$ in the respective games $\Game_r^M(\StrA_1,\StrB_1)$ and
  $\Game_r^M(\StrA_2,\StrB_2)$.

  Since $A_1$ and $A_2$ as well as $B_1$ and $B_2$ are disjoint, we
  have $\card{S} = \card{S_1} + \card{S_2}$ and $\card{D} = \card{D_1}
  + \card{D_2}$. Furthermore, $\card{S_1} \equiv \card{D_1} \pmod M$
  and $\card{S_2} \equiv \card{D_2} \pmod M$ as the sets $D_1$ and
  $D_2$ are chosen according to Duplicator's winning strategies in the
  games on $\StrA_1$ and $\StrB_1$, and $\StrA_2$ and $\StrB_2$,
  respectively. Since $\equiv\pmod M$ is a congruence relation with
  respect to addition, we have that $\card{S} \equiv \card{D} \pmod
  M$. It is easily verified that the sets and elements chosen
  according to this strategy indeed define a twofold partial
  isomorphism between $\StrA$ and $\StrB$.
\end{proof}

As a direct corollary we obtain the following result that will be
used in the inductive step in the forthcoming proofs.

\begin{corollary}\label{cor:cmso-disjoint-union-exp}
  Let $\StrA_1, \StrA_2, \StrB_1,$ and $\StrB_2$ be $\tau$-structures,
  such that $\StrA_1 \equiv_r^M \StrB_1$ and $\StrA_2 \equiv_r^M
  \StrB_2$. Then $(\StrA_1 \dunion \StrA_2,A_1) \equiv_r^M (\StrB_1
  \dunion \StrB_2, B_1)$.
\end{corollary}
\begin{proof}
  We consider the following $\tau \dunion \{P\}$-expansions of the
  given structures: $\StrA'_1 := (\StrA_1,A_1)$, $\StrB'_1 :=
  (\StrB_1,B_1)$, $\StrA'_2 := (\StrA_2,\emptyset)$, and $\StrB'_2 :=
  (\StrB_2,\emptyset)$. It is immediate that
  \begin{enumerate}[(i)]
    \setlength{\itemsep}{1ex plus0.5ex minus0.5ex}
  \item $\StrA_1 \equiv_r^M \StrB_1$ implies $(\StrA_1,A_1) \equiv_r^M
    (\StrB_1,B_1)$, and
  \item $\StrA_2 \equiv_r^M \StrB_2$ implies $(\StrA_2,\emptyset)
    \equiv_r^M (\StrB_2,\emptyset)$
  \end{enumerate}
  since Duplicator can obviously win the respective \EF games on the
  expanded structures using the same strategies as in the games
  proving the equivalences on the left-hand side. The claim follows by
  applying the previous lemma to the $\tau \dunion \{P\}$-expansions.
\end{proof}

It is well known that \MSO exhibits a certain weakness regarding the
ability to specify cardinality constraints on sets, \ie, structures
over an empty vocabulary. A proof of this fact using \EF games can be
found in \cite{Libkin:FMT}. By adapting this proof, we show that this
is still the case for \CMSO.

\begin{lemma}\label{lem:cmso-set-equiv}
  Let $\StrA$ and $\StrB$ be $\emptyset$-structures,
  $M\finsubseteq\setN^+$, and $r\in\setN$. Then $\StrA \equiv_r^M
  \StrB$ if $|A|,|B| \geq (2^{r+1}-4)\lcm(M)$ and $|A| \equiv |B|
  \pmod M$.
\end{lemma}
\begin{proof}
  We prove by induction on the number of rounds that Duplicator wins
  the (mod $M$) $r$-round \EF game $\Game_r^M(\StrA,\StrB)$. For $r=0$
  and $r=1$ the claim is obviously true. Let $r>1$, assume that the
  claim holds for $r-1$, and consider the first move of the $r$-round
  game. We assume that Spoiler makes his move in $\StrA$ since the
  reasoning in the other case is completely symmetric.

  If Spoiler makes a set move $S\subseteq A$, we consider the
  following cases:
  \begin{enumerate}[(1)]
  \item $|S| < (2^r-4)\cdot\lcm(M)$ (or $|A-S| <
    (2^r-4)\cdot\lcm(M)$). Then Duplicator selects a set $D \subseteq
    B$ such that $|D| = |S|$ (or $|B-D|=|A-S|$), and hence $S \isom D$
    and $A-S \equiv_{r-1}^M B-D$ (or $A-S \isom B-D$ and $S
    \equiv_{r-1}^M D$).
  \item $|S|,|A-S| \geq (2^r-4)\cdot\lcm(M)$. Then Duplicator selects
    a set $D \subseteq B$ such that $|D| \equiv |S| \pmod M$ and
    $|D|,|B-D| \geq (2^r-2)\cdot\lcm(M)$. In fact, she chooses for $D$
    half of the elements and chooses $\ell < \lcm(M)$ additional ones
    to fulfil the cardinality constraints $|D| \equiv |S| \pmod M$.
    Then, for the set $B-D$ of non-selected elements, we have
    \begin{align*}%
      |B-D| &\geq \frac{1}{2}\big((2^{r+1}-4)\lcm(M)\big) - \ell \geq
      (2^r-2)\lcm(M) - \lcm(M)\\ &\geq (2^r - 4)\lcm(M)
    \end{align*}
    for all $\ell$ satisfying $0 \leq \ell < \lcm(M)$. Since $|D| =
    |B-D| + 2\ell$, obviously $|D| \geq (2^r-4)\lcm(M)$ as well.
  \end{enumerate}
  Thus, in both cases, by the induction hypothesis we get $S
  \equiv_{r-1}^M D$ and $A-S \equiv_{r-1}^M B-D$. Hence, by
  Corollary~\ref{cor:cmso-disjoint-union-exp} $(A,S) \equiv_{r-1}^M
  (B,D)$, \ie, Duplicator has a winning strategy in the remaining
  $(r-1)$-round game from position $(S,D)$.

  If Spoiler makes a point move $s \in A$, Duplicator answers by
  choosing an arbitrary el\-e\-ment $d \in B$. Similar to Case~1
  above, we observe that $(\{s\},s) \isom (\{d\},d\,)$ and $A-\{s\}
  \equiv_{r-1}^M B-\{d\}$ by the induction hypothesis. Thus, by
  Lemma~\ref{lem:cmso-disjoint-union}, $(A,s) \equiv_{r-1}^M (B,d)$
  implying that Duplicator has a winning strategy for the remaining
  $r-1$ rounds from position $(s,d)$.
\end{proof}

\section{The Separating Example}

We will first give a brief description of our example showing that
\oiMSO is strictly more expressive than \CMSO. We consider a property
of two-dimensional grids, namely that the vertical dimension divides
the horizontal dimension. This property is easily definable in \MSO
for grids that are given as directed graphs with two edge relations,
one for the horizontal edges pointing rightwards, and one for the
vertical edges pointing upwards, by defining a new relation of
diagonal edges combining one step rightwards and one step upwards
wrapping around from the top border to the
bottom border but not from the right to the left border. Note that
there is a path following those diagonal edges starting from the
bottom-left corner of the grid and ending in the top-right corner if,
and only if, the vertical dimension divides the horizontal dimension
of the grid. Thus, for our purposes, we have to weaken the structure
in the sense that we hide information that remains accessible to
\oiMSO-formulae but not to \CMSO formulae.

An appropriate loss of information is achieved by replacing the two
edge relations with their reflexive symmetric transitive closure, \ie,
we consider grids as structures with two equivalence relations which
provide a notion of \emph{rows} and \emph{columns} of the grid.
Obviously, notions like corner and border vertices as well as the
notion of an order on the rows and columns that were important for the
MSO-definition of the divisibility property are lost, but
clearly, all these notions can be regained in presence of an order.
First, the order allows us to uniquely define an element (\eg the
$<$-least element) to be the bottom-left corner of the grid, and
second, the order induces successor relations on the set of columns
and the set of rows, from which both horizontal and vertical successor
vertices of any vertex can be deduced. Since the divisibility property
is obviously invariant with respect to the ordering of the rows or
columns, this allows for expressing it in \oiMSO. In the course of
this section we will develop the arguments showing that \CMSO fails to
express this property on the following class of grid-like structures.

\begin{definition}
  A \emph{cliquey $(k,\ell)$-grid} is a $\{\sim_h,\sim_v\}$-structure
  that is isomorphic to $\StrG_{k\ell} := (\{0,\dots,k-1\}\times
  \{0,\dots,\ell-1\}, \sim_h, \sim_v)$, where
  \begin{align*}
    \sim_h\ & := \{((x,y),(x',y')) : x=x'\} \text{ and}\\
    \sim_v\ & := \{((x,y),(x',y')) : y=y'\}\,,
  \end{align*}
  \ie, $\sim_h$ consists of exactly $k$ equivalence classes (called
  \emph{rows}), each containing $\ell$ elements, and $\sim_v$ consists
  of exactly $\ell$ equivalence classes (called \emph{columns}), each
  containing $k$ elements, such that every equivalence class of
  $\sim_h$ intersects every equivalence class of $\sim_v$ in exactly
  one element and vice versa.

  A \emph{horizontally coloured cliquey $(k,\ell)$-grid}, denoted
  $\ColG_{k\ell}$, is the expansion of the $\{\sim_v\}$-reduct of the
  cliquey grid $\StrG_{k\ell}$ by unary predicates
  $\{P_1,\dots,P_k\}$, where the information of~$\sim_h$ is retained
  in the $k$ new predicates (in the following referred to as
  \emph{colours}) such that each set $P_i$ corresponds to exactly one
  former equivalence class.
\end{definition}

Note that the same class of grid-like structures has already been used
by Otto in a proof showing that the number of monadic second-order
quantifiers gives rise to a strict hierarchy over finite
structures~\cite{Ot95c}.

The class is first-order definable by a sentence~$\psi_\text{grid}$
stating that
\begin{itemize}
  \setlength{\itemsep}{1ex plus0.5ex minus0.5ex}
  \item $\sim_v$ and $\sim_h$ are equivalence relations, and
  \item every pair consisting of one equivalence class of $\sim_h$ and
    $\sim_v$ each has exactly one element in common
\end{itemize}
as these properties are sufficient to enforce the desired grid-like
structure. Note that even the second property is first-order definable
since every equivalence class is uniquely determined by each of its
elements.

The following two lemmata justify the introduction of the notion of
horizontally coloured cliquey grids for use in the forthcoming proofs.

\begin{lemma}
  \label{lem:disjoint-grid-comp}
  Let $\ColG_{k\ell_1}$, $\ColG_{k\ell_2}$, $\ColG_{k\ell'_1}$, and
  $\ColG_{k\ell'_2}$ be horizontally coloured cliquey grids such that
  $\ColG_{k\ell_1} \equiv_r^M \ColG_{k\ell'_1}$ and $\ColG_{k\ell_2}
  \equiv_r^M \ColG_{k\ell'_2}$. Then $\ColG_{k,\ell_1+\ell_2} \equiv_r^M
  \ColG_{k,\ell'_1+\ell'_2}$.
\end{lemma}
\begin{proof}
  Note that, since there are no horizontal edges in horizontally
  coloured cliquey grids and the vertical dimension of all grids is
  $k$, $\ColG_{k,\ell_1+\ell_2}$ is the disjoint union of the two
  smaller horizontally coloured cliquey grids $\ColG_{k\ell_1}$ and
  $\ColG_{k\ell_2}$, and of course, the same holds for
  $\ColG_{k,\ell'_1+\ell'_2}$. Thus, the claim follows by
  Lemma~\ref{lem:cmso-disjoint-union}.
\end{proof}

\begin{lemma}\label{lem:col-to-noncol}
  Let $\ColG_{k\ell} \equiv_r^M \ColG_{k\ell'}$. Then $\StrG_{k\ell}
  \equiv_r^M \StrG_{k\ell'}$.
\end{lemma}
\begin{proof}
  For each fixed horizontal dimension $k$, there exists a
  one-dimensional quantifier-free interpretation of a cliquey grid in
  its respective horizontally coloured counterpart since we can define
  the horizontal equivalence relation $\sim_h$ in terms of the colours
  as follows:
  \[ x \sim_h y\ \equiv\ \biglor_{i=1}^k P_i x \land P_i
  y\,.\]
  \vskip -5ex
\end{proof}

Actually, the argument implies that Duplicator wins a game on cliquey
grids using the same strategy that is winning in the corresponding
game on coloured grids since a strategy preserving the colours of
selected elements especially preserves the equivalence relation
$\sim_h$.

Before stating the main lemma, we will first prove a combinatorial
result which will later help Duplicator in synthesising her winning
strategy and introduce the following weakened notion of equality
between numbers.

\begin{definition}
  Two numbers $a, b \in \setN$ are called \emph{threshold $t$ equal
    \emph{(}mod~$M$\emph{)}}, denoted $a =^M_t b$, if
  \begin{enumerate}[(i)]
    \setlength{\itemsep}{1ex plus0.5ex minus0.5ex}
  \item $a = b$ or
  \item $a,b \geq t$ and $a \equiv b \pmod M$.
  \end{enumerate}
  Intuitively, $a =^M_t b$ means that the numbers are equal if they
  are small, or that they are at least congruent modulo all $m \in M$
  if they are both at least as large as the threshold~$t$.
\end{definition}

\begin{lemma}
  \label{lem:combinatorics}
  For every $p, t \in \setN$, and $M \finsubseteq \setN^+$, we can
  choose an arbitrary $T \geq p \cdot (t + \lcm(M) - 1)$ such that for
  all sets $A$ and $B$ with $\card{A} =^M_T \card{B}$ and for every
  equivalence relation $\eqrel_A$ on $A$ of index at most~$p$ there
  exists an equivalence relation $\eqrel_B$ on $B$ and a bijection $g
  \colon \quot{A}{\eqrel_A} \to \quot{B}{\eqrel_B}$ satisfying
  $\card{\{a' \in A : a \eqrel_A a'\}} =^M_t \card{g(\{a' \in A : a
    \eqrel_A a'\})}$ for all $a\in A$.
\end{lemma}

\begin{proof}
  We let $\{\seq[p']a\}$, where $p' \leq p$ denotes the index of
  $\eqrel_A$, be the set of class representatives of
  $\quot{A}{\eqrel_A}$, and we let $[a]_{\eqrel_A} := \{ a'\in A : a'
  \eqrel_A a\}$ denote the equivalence class of $a$ in $A$. Note that
  we will usually omit the subscript $\eqrel_A$ if it is clear from
  the context and instead reserve the letters $a$ and $b$ for elements
  denoting equivalence classes in $A$ and $B$, respectively.
  Furthermore, a set will be called \emph{small} in the following if
  it contains less than $t$ elements and \emph{large} otherwise.

  The equivalence relation $\eqrel_B$ on $B$ is constructed by
  partitioning the set into $p'$ disjoint non-empty subsets
  $\{\seq[p']{B}\}$ as follows. If $\card{A} = \card{B}$, for each
  class $[a_i]$, we choose a set~$B_i$ with exactly $\card{[a_i]}$
  many elements. If $\card{A}, \card{B} \geq T$, we have to
  distinguish between the treatment of small and large classes. Since
  $\card{A} \geq T \geq p \cdot (t + \lcm(M) - 1)$, $\lcm(M) \geq 1$,
  and the index of $\eqrel_A$ is at most $p$, at least one of the
  equivalence classes contains at least $t$ elements, \ie, it is
  large, and without loss of generality, it is assumed that this is
  the case for $[a_1]$. For each small class~$[a_i]$, we choose a
  set~$B_i$ with exactly $\card{[a_i]}$ many elements.  If~$[a_i]$ is
  large, we choose a set~$B_i$ containing $t + \ell$ many elements
  where~$\ell$ is the smallest non-negative integer such that
  $\card{[a_i]} \equiv \card{B_i} \pmod M$. The number of elements
  selected according to these rules is at most $p \cdot (t + \lcm(M) -
  1) \leq T \leq \card{B}$. Since $[a_1]$ is large by assumption, any
  possibly remaining elements in $B$, that have not been assigned to
  one of the subsets $\seq[p']{B}$ yet, can be safely added to $B_1$
  without violating the condition that $\card{[a_1]} \equiv \card{B_1}
  \pmod M$.

  This partitioning uniquely defines the equivalence relation
  $\eqrel_B := \bigcup_{i=1}^{p'} (B_i \times B_i)$ on~$B$. By
  selecting an arbitrary element of each $B_i$ we get a set of class
  representatives $\{\seq[p']b\}$ which directly yields the bijection
  $g \colon [a_i] \mapsto [b_i]$ for all $1 \leq i\leq p'$ satisfying
  $\card{[a]} =^M_t \card{g([a])}$ for all $a\in A$ by construction.
\end{proof}

The following lemma extends the results on \CMSO-equivalence of
\emph{large enough sets} to \emph{large enough grids} by giving a
sufficient condition on the sizes of two grids for the existence of a
winning strategy for Duplicator in an $r$-round (mod $M$) game on the
two structures. Due to the inductive nature of the proof that
involves, in each step, a construction of equivalence classes as in
the above lemma, we need as a criterion for the size, for fixed $p \in
\setN$ and $M \finsubseteq \setN^+$, a function $f_{p,M} : \setN \to
\setN$ such that, for all $r \in \setN^+$ and $t = f_{p,M}(r-1)$, we
can choose $T = f_{p,M}(r)$ in the previous lemma. One function
satisfying, for all $r \in \setN^+$, the inequality $f_{p,M}(r) \geq p
\cdot (f_{p,M}(r-1) + \lcm(M) - 1)$ derived from the condition imposed
on~$T$ is $f_{p,M}(r) = 2\cdot(p^r-1)\cdot\lcm(M)$.

\begin{lemma}\label{lem:cmso-grid-equiv}
  Let $M \finsubseteq \setN^+$, $r \in \setN$ and $k > 1$ be fixed.
  Then for $f(r) := f_{2^k,M}(r) = (2^{kr+1}-2)\lcm(M)$, as given
  above, $\StrG_{k\ell_1} \equiv_r^M \StrG_{k\ell_2}$ if $\ell_1
  =^M_{f(r)} \ell_2$.
\end{lemma}
\begin{proof}
  As motivated by Lemma~\ref{lem:col-to-noncol}, we consider the
  $r$-round (mod $M$) \EF game on the corresponding horizontally
  coloured cliquey grids $\ColG_{k\ell_1}$ and $\ColG_{k\ell_2}$, and
  we show by induction on the number of rounds that Duplicator has a
  winning strategy in this game.

  Intuitively, the proof proceeds as follows. Spoiler's set move
  induces an equivalence relation on the set of columns forming the
  grid he plays in, and the previous lemma implies that Duplicator is
  able to construct an equivalence relation on the columns of the
  other grid which is similar in the sense that corresponding
  equivalence classes satisfy certain cardinality constraints. Since
  the grids can be regarded as disjoint unions of these equivalence
  classes, we can argue by induction that corresponding subparts of
  the two grids, being similar enough, cannot be distinguished during
  the remaining $r-1$ rounds of the game.

  The case where $\ell_1 = \ell_2$ is trivial since grids of the same
  dimensions are isomorphic. Thus, we assume in the following that
  $\ell_1,\ell_2 \geq f(r)$ and $\ell_1 \equiv \ell_2 \pmod M$. The
  claim is obviously true for $r=0$, hence we assume that it holds for
  $r-1$ and proceed with the inductive step. As before, we assume
  without loss of generality that Spoiler makes his moves in
  $\StrG_{k\ell_1}$ since the other case is symmetric.

  A \emph{coloured $k$-column} is a
  $\{\sim_v,P_1,\dots,P_k\}$-structure isomorphic to $\ColC_k :=
  \ColG_{k,1}$, such that a coloured grid can be regarded as a
  disjoint union of columns. Given a subset $S$ of vertices of a grid
  and one of its coloured $k$-columns $\StrC$ with universe $C$, the
  \emph{colour-type of $\StrC$ induced by $S$} is defined as the
  isomorphism type of the expansion $(\StrC,S \cap C)$ denoted by
  $\isotype(\StrC,S)$.  Given a set~$\mathcal F$ of $k$-columns, each
  subset~$S$ of all of their vertices gives rise to an equivalence
  relation $\eqrel_S$ on $\mathcal F$ by virtue of $\StrC_1 \eqrel_S
  \StrC_2$ if, and only if, $\isotype(\StrC_1,S) =
  \isotype(\StrC_2,S)$. Note that the index of $\eqrel_S$ is at most
  $2^k$.

  Assume, Spoiler performs a set move and chooses a subset $S$ in
  $\ColG_{k\ell_1} = \StrC_1 \dunion \cdots \dunion \StrC_{\ell_1}$. As
  described above, $S$ induces an equivalence relation $\eqrel_S$ with
  at most $2^k$ equivalence classes on the set $\mathcal F =
  \{\StrC_1,\dots,\StrC_{\ell_1}\}$ of columns forming the grid. For
  $p = 2^k$, $t = f(r-1)$ and $M$ as given, by the previous lemma,
  there is an equivalence relation $\eqrel'_S$ on the set $\mathcal F'
  = \{\StrC'_1,\dots,\StrC'_{\ell_2}\}$ of columns on the Duplicator's
  grid $\ColG_{k\ell_2}$ since $\ell_1,\ell_2 \geq f(r)$. Furthermore,
  there is a bijection $g$ mapping equivalence classes of columns in
  one grid to the other.

  Given that the index of both $\eqrel_S$ and $\eqrel'_S$ is $p' \leq
  p=2^k$, we can assume $\{\seq[p']{\StrC}\}$ and
  $\{\seq[p']{\StrC'}\}$ to be the sets of class representatives of
  $\eqrel_S$ and $\eqrel'_S$, respectively. Duplicator now selects the
  unique set $D$ of elements such that $\isotype(\StrC,S) =
  \isotype(\StrC',D)$ for all $1 \leq i\leq p'$, $\StrC \in [\StrC_i]$
  and $\StrC' \in g([\StrC_i])$.

  For each $1 \leq i\leq p'$, we let $\indStr{\StrC_i} :=
  \ColG_{k\ell_1}\!\restr{[\StrC_i]}$ and $\indStr{\StrC'_i} :=
  \ColG_{k\ell_2}\!\restr{[\StrC'_i]}$ denote the substructures of the
  grids $\ColG_{k\ell_1}$ and $\ColG_{k\ell_2}$ induced by the sets of
  columns $[\StrC_i]$ and $[\StrC'_i]$, respectively. By construction,
  we have $\card{[\StrC_{i}]} =^M_{f(r-1)} \card{[\StrC'_{i}]}$ for
  all $i$. Thus, depending on whether $[\StrC_{i}]$ (and hence
  $[\StrC'_{i}]$) are small or large with respect to the threshold
  $f(r-1)$, either $\indStr{\StrC_{i}} \isom
  \indStr{\StrC'_{i}}$ or $\indStr{\StrC_{i}} \equiv_{r-1}^M
  \indStr{\StrC'_{i}}$ by the induction hypothesis.  Since $S$ and
  $D$ induce the same colour-types on the columns in $[\StrC_{i}]$ and
  $[\StrC'_{i}]$, respectively, we have
  \[\big(\indStr{\StrC_{i}},S \cap
  \univ{\indStr{\StrC_{i}}}\big) \equiv_{r-1}^M
  \big(\indStr{\StrC'_{i}},D \cap
  \univ{\indStr{\StrC'_{i}}}\big)\]
  for all $i$, where $\univ{\cdot}$ denotes the
  universe of the respective structure.  Thus, iterating
  Lemma~\ref{lem:cmso-disjoint-union} yields that Duplicator has a
  winning strategy in the remaining rounds of the game
  $\Game_{r-1}^M(\ColG_{k\ell_1},\ColG_{k\ell_2})$ from position
  $(S,D)$.

  If Spoiler makes a point move $s$, say in column $\StrC_1$ of the
  grid $\ColG_{k\ell_1}$, Duplicator picks an arbitrary element $d$ of
  the same colour in her grid, say in column $\StrC'_1$. As the
  substructures consisting of just the columns containing the chosen
  elements are isomorphic, \ie, $\big(\StrC_{1},s\big) \isom
  \big(\StrC'_{1},d\big)$, and by the induction hypothesis we have
  $\StrC_{2} \dunion \cdots \dunion \StrC_{\ell_1} \equiv^M_{r-1}
  \StrC'_{2}\dunion\cdots\dunion\StrC'_{\ell_2}$, Duplicator can win
  the remaining $(r-1)$-round game from position $(s,d)$ by
  Lemma~\ref{lem:cmso-disjoint-union}.
\end{proof}

Now we have the necessary tools available to prove the main theorem.

\begin{theorem}
  $\CMSO \subsetneq \oiMSO$.
\end{theorem}
\begin{proof}
  We show that the class $\ClsC := \{\,\StrG_{k\ell}\ \colon\ k |
  \ell\,\}$ is not definable in \CMSO but order-invariantly definable in
  \MSO by the sentence $\psi_\text{grid} \land \phi$, where
  \begin{align*}
    \phi &= \exists \min\exists c \left(\begin{aligned}
        & \forall x (\min \leq x) \land \neg\exists z (E_h(c,z) \lor E_v(c,z))\\
        \land\ & \forall T \big(\forall x\forall y (Tx \land
        \phi_\text{diag}(x,y) \limp Ty) \land T\min{} \limp Tc\big)
      \end{aligned}\right)\ ,
  \end{align*}
  and
  \begin{align*}
    \phi_\text{diag}(x,y) &= \big(\exists z (E_v(x,z) \land E_h(z,y))\big)\\
    &\qquad \lor \big(\neg\exists z E_v(x,z) \land \exists z (z \sim_h
    \min{}
    \land z \sim_v x \land E_h(z,y))\big)\ ,\\
    E_h(x,y) &= x \sim_h y \land \exists x_0\exists y_0
    \left(\begin{aligned}
      & x_0 \sim_h \min{} \land y_0 \sim_h \min{}\\
      \land\ & x \sim_v x_0 \land y \sim_v y_0 \land x_0 < y_0\\
      \land\ & \forall z_0 ( z_0
      \sim_h \min{} \limp z_0 \leq x_0 \lor z_0 \geq y_0)\quad ,
      \!\!\!\!\!\!\!
    \end{aligned}\right)\\
    E_v(x,y) &= x \sim_v y \land \exists x_0\exists y_0
    \left(\begin{aligned}
      & x_0 \sim_v \min{} \land y_0 \sim_v \min{}\\
      \land\ & x \sim_h x_0 \land y \sim_h y_0 \land x_0 < y_0\\
      \land\ & \forall z_0 ( z_0
      \sim_v \min{} \limp z_0 \leq x_0 \lor z_0 \geq y_0)\quad .
      \!\!\!\!\!\!\!
    \end{aligned}\right)
  \end{align*}
  As hinted above, the horizontal and vertical edge relations ($E_h$
  and $E_v$, respectively) are defined using the successor relation
  which is induced by an arbitrary ordering on the row (and column)
  containing the minimal element ($\min$) which itself serves as the
  lower left corner of the grid. $\phi_\text{diag}$ defines diagonal
  steps through the grid that wrap around from the top to the bottom
  row. Finally, $\phi$ states that the pair consisting of the lower
  left corner ($\min$) and the upper right corner ($c$) of the grid is
  contained in the transitive closure of $\phi_\text{diag}$.
  Obviously, there is such a sawtooth-shaped path starting at $\min$
  and ending exactly in the upper right corner if, and only if, $k |
  \ell$.

  The second step consists in showing that $\ClsC$ is not definable in
  \CMSO. Towards this goal, we show that for any choice of $r \in
  \setN$ and $M \finsubseteq \setN^+$, we can find $k,\ell_1,\ell_2
  \in \setN$, such that $\StrG_{k\ell_1} \in \ClsC$, $\StrG_{k\ell_2}
  \not\in \ClsC$, and $\StrG_{k\ell_1} \equiv_r^M \StrG_{k\ell_2}$
  which contradicts the \CMSO-definability of~$\ClsC$.

  Let $r \in \setN$ and $M \finsubseteq \setN^+$ be fixed. We choose
  $s \geq r+1$ such that $2^s \notdiv \lcm(M)$. Let $k = 2^s$, $\ell_1
  = 2^{kr+1}\lcm(M)$, and $\ell_2 = \ell_1 + \lcm(M)$.  Obviously,
  $\ell_1$ and $\ell_2$ satisfy the conditions of
  Lemma~\ref{lem:cmso-grid-equiv}, and thus $\StrG_{k\ell_1}
  \equiv_r^M \StrG_{k\ell_2}$.

  Furthermore, $\ell_1 = k \cdot 2^{2^s\cdot r-s+1}\lcm(M)$, hence $k
  \mid \ell_1$ and $\StrG_{k\ell_1} \in \ClsC$. On the other hand, $k
  \notdiv \ell_2 = \ell_1+\lcm(M)$ by the choice of $s$, thus
  $\StrG_{k\ell_2} \not\in \ClsC$.
\end{proof}

\vskip -0.5cm
\section{Conclusion}

We have provided a characterisation of the expressiveness of \CMSO in
terms of an \EF game that naturally extends the known game capturing
\MSO-de\-fin\-abil\-i\-ty, and we have presented a class of structures that
are shown, using the proposed game characterisation, to be undefinable
by a \CMSO-sentence yet being definable by an order-invariant
\MSO-sentence. This establishes that order-invariant \MSO is strictly
more expressive than counting \MSO in the finite. Modifying the
separating example by considering a variant of cliquey grids where the
two separate equivalence relations are unified into a single binary
relation and considering, \eg, the class of such grids where the
horizontal dimension exactly matches the vertical dimension, we can
also confirm Courcelle's original conjecture.

\begin{corollary}
  \CMSO-definability is strictly weaker than \oiMSO-definability for
  general graphs.
\end{corollary}

The separating query being essentially a transitive closure query,
\ie, the only place where monadic second-order quantification is used
is in the definition of the transitive closure of a binary relation,
we can conclude that the same class of structures yields a separation
of $\mathInvLogic{(D)TC^1}{<}$ from $\text{(D)TC}^1$ (the extension of
\FO by a (deterministic) transitive closure operator on binary
relations) and even from $\text{(D)TC}^1$ extended with
modulo-counting predicates since $\text{(D)TC}^1 \subseteq \MSO$.
Finding separating examples concerning higher arity (D)TC or even full
(D)TC requires further investigation since, in general, $\MSO
\subsetneq \DTC^2$.

Following an opposite line of research, it would be interesting to
identify further classes of graphs, besides classes of graphs of
bounded tree-width, on which $\oiMSO$ is no more expressive than
\CMSO.

\bibliographystyle{alpha}

\vskip -0.6cm

\end{document}